\documentclass[
superscriptaddress,
nofootinbib,
twocolumn,
 amsmath,amssymb,
 aps,
pra,
]{revtex4-1}

\usepackage{dcolumn}
\usepackage{bm}
\usepackage{epsfig}
\usepackage{graphicx}
\usepackage{latexsym}
\usepackage{amsfonts}
\usepackage{setspace}
\usepackage{graphicx}
\usepackage{amsmath}
\usepackage{verbatim}
\usepackage{color}
\usepackage{SIunits}
\usepackage{hyperref}

\newcommand{\be}{\begin{equation}}
\newcommand{\ee}{\end{equation}}
\newcommand{\ba}{\begin{array}}
\newcommand{\ea}{\end{array}}
\newcommand{\bqa}{\begin{eqnarray}}
\newcommand{\eqa}{\end{eqnarray}}

\begin{document}

\title{Phonon routing in integrated optomechanical cavity-waveguide systems}

\author{Kejie Fang} 
\author{Matthew H. Matheny}
\author{Xingsheng Luan}
\affiliation{Kavli Nanoscience Institute and Thomas J. Watson, Sr., Laboratory of Applied Physics, California Institute of Technology, Pasadena, California 91125, USA}
\affiliation{Institute for Quantum Information and Matter, California Institute of Technology, Pasadena, California 91125, USA}
\author{Oskar Painter}
\email{opainter@caltech.edu}
\affiliation{Kavli Nanoscience Institute and Thomas J. Watson, Sr., Laboratory of Applied Physics, California Institute of Technology, Pasadena, California 91125, USA}
\affiliation{Institute for Quantum Information and Matter, California Institute of Technology, Pasadena, California 91125, USA}

\begin{abstract} 
The mechanical properties of light have found widespread use in the manipulation of gas-phase atoms and ions, helping create new states of matter and realize complex quantum interactions.  The field of cavity-optomechanics strives to scale this interaction to much larger, even human-sized mechanical objects.  Going beyond the canonical Fabry-Perot cavity with a movable mirror, here we explore a new paradigm in which multiple cavity-optomechanical elements are wired together to form optomechanical circuits.  Using a pair of optomechanical cavities coupled together via a phonon waveguide we demonstrate a tunable delay and filter for microwave-over-optical signal processing.  In addition, we realize a tight-binding form of mechanical coupling between distant optomechanical cavities, leading to direct phonon exchange without dissipation in the waveguide.  These measurements indicate the feasibility of phonon-routing based information processing in optomechanical crystal circuitry, and further, to the possibility of realizing topological phases of photons and phonons in optomechanical cavity lattices.
\end{abstract}

\maketitle

\section{Introduction} 
\label{sec:intro}
Microscopic optical and microwave cavities, with their wavelength or even sub-wavelength mode size, have recently been used to greatly enhance the radiation pressure interaction between electromagnetic waves trapped in the cavity and mechanical vibrations of the cavity walls~\cite{RMP}.  Technical advances in our ability to fabricate structures of ever smaller size and higher quality has led to the demonstration of a number of new phenomena, including laser cooling of mechanical resonators into their quantum ground state~\cite{cool1,cool2}, optomechanically-induced electromagnetic transparency~\cite{eit1,eit2,eit3}, and microwave signal processing~\cite{microwave1,microwave2}.  To date these experiments have largely involved single cavity elements.  A new paradigm, enabled by chip-scale cavity-optomechanics, involves the integration of multiple optomechanical cavities together to realize extended photon-phonon excitations for classical and quantum information processing applications~\cite{ophwg1,ophwg2,ophwg3}. Among the different types of optomechanical systems currently being studied, the device architecture based on thin-film optomechanical crystals~\cite{phoxonic1,phoxonic2,omc,shielding,phoxonic3} is a particularly promising integration platform given the large attainable radiation-pressure coupling and the ability to create photonic and phononic band-gap waveguides.  

In an integrated optomechanical network composed of cavities and waveguides, propagating phonons can be routed and stored among different nodes as controlled by driving laser fields, providing various information processing functions - such as buffering, delay, and filtering - in a chip-scale footprint due to the slow velocity of acoustic waves~\cite{fiber_delay_line1,onchip_true_delay_line1,onchip_true_delay_line2}.  As an initial step towards realizing more complex multi-element cavity-optomechanical systems, here we explore the optical excitation and routing of GHz phonons in a small optomechanical circuit consisting of two optomechanical crystal cavities connected by a dispersion-engineered phonon waveguide.  Pulsed and continuous-wave measurements are first used to characterize the coupling efficiency and phonon propagation properties of the system.  Utilizing separate optical driving fields for each optomechanical cavity, we use this system to demonstrate an optically-tunable microwave delay line in which microwave-over-optical signals sent in one cavity port are efficiently converted into outgoing microwave-over-optical signals on the other cavity port.  

Owing to the continuum of modes supported by the phonon waveguide, a tight-binding type of interaction between distant cavities may also be realized.  Such ideas have recently been explored utilizing photonic waveguides, both in the optical domain with photonic crystal cavities~\cite{noda} and in the microwave domain with superconducting qubits~\cite{wallraff}.  Here we use a phonon waveguide to strongly couple two distant optomechanical cavities, leading to an exchange-type phonon routing.  The hybridized mechanical cavity modes, while spatially extended, nevertheless have strong optomechanical coupling with both localized optical cavity modes which are not hybridized. Such a feature can be used to create mechanical transducers capable of bridging two separated physical systems~\cite{lukin}, such as in quantum wavelength conversion between optical and superconducting microwave systems~\cite{lambda1,lambda2,lambda3,lambda4} and coupling between solid state spins~\cite{spin1,spin2,spin3}.  Scaling such strong coupling to arrays of cavity elements may also enable novel optomechanical metamaterials, where in combination with a spatially distributed phase pattern of the optical driving field~\cite{fang}, topological phases for photons and phonons are expected to arise~\cite{manybody1,manybody2}. 


\begin{figure*}[btp]
\begin{center}
\includegraphics[width=2\columnwidth]{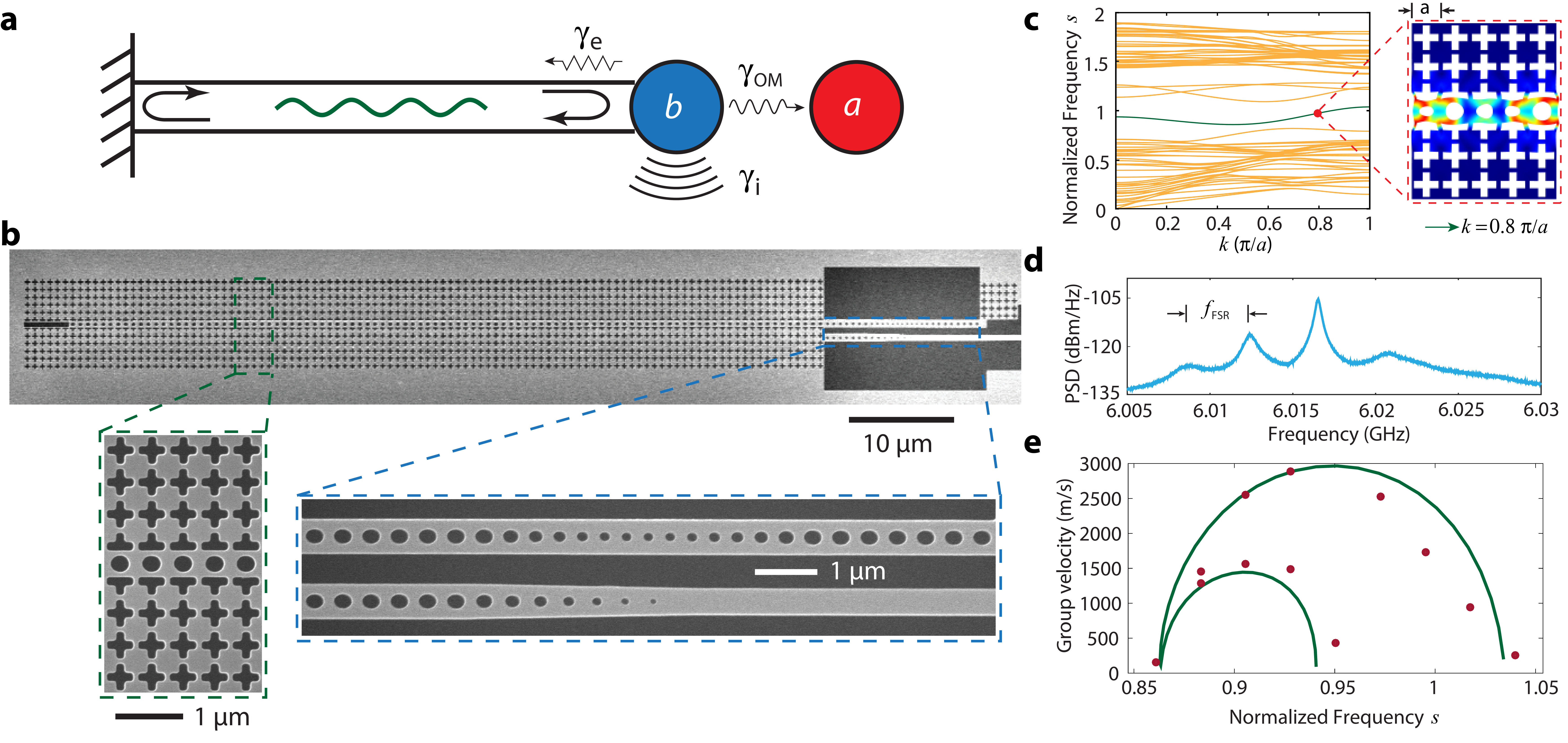}
\caption{\textbf{GHz dispersion-engineerable phonon waveguide.} \textbf{a}, A schematic diagram of the single cavity-waveguide device. $a$ and $b$ represent the optical and mechanical cavity modes. The mechanical cavity mode couples to the phonon waveguide at rate $\gamma_e$ and has an optically induced decay rate $\gamma_{\rm OM}$. \textbf{b}, Scanning electron microscope (SEM) image of a silicon optomechanical cavity-waveguide device based on the scheme of \textbf{a}. \textbf{c}, Theoretical band structure of the phonon waveguide employed in this work. Green curve is the waveguide breathing mode and yellow curves are the bands of the phononic shielding. Inset on the right is the modal profile of the waveguide breathing mode corresponding to wavevector $k=0.8\frac{\pi}{a}$. \textbf{d}, Optically transduced mechanical power spectral density (PSD) of a typical device showing a few hybridized cavity-waveguide resonances. \textbf{e}, Group velocity as a function of frequency across the breathing mode waveguide band. Green is the theoretical curve and red dots are the measured data.}
\label{fig1}
\end{center}
\end{figure*}

\section{Phonon Waveguide Characterization}
\label{sec:phonon_waveguide}
While integrated phonon waveguides with frequency bands of kHz to a few hundred MHz have been previously studied~\cite{wg2,wg3,wg4}, here we explore the properties of thin-film, dispersion engineered phonon waveguides in the GHz frequency range.  A schematic diagram and SEM image of a cavity-waveguide device used to study such phonon waveguide properties are shown in Fig.~\ref{fig1}a and Fig.~\ref{fig1}b, respectively. The device is fabricated from the $220$~nm silicon device layer of a silicon-on-insulator wafer (see App.~\ref{app:methods}), and consists of several parts: an optomechanical cavity with co-localized optical and acoustic resonances, an optical coupling waveguide, and a long phonon waveguide section which is end-fire coupled to the cavity.  The optomechanical cavity is realized in a photonic crystal nanobeam which supports an optical resonance at $\lambda_c \approx 1550$~nm in the telecom wavelength band that is strongly coupled through the elasto-optic radiation pressure effect to a ``breathing mode'' mechanical resonance of the beam at a frequency of $\omega_m/2\pi \approx 6$~GHz. One end of the nanobeam is terminated by a two dimensional phononic crystal mirror with a full phononic bandgap in the relevant frequency range~\cite{shielding}; the other end is connected to the phonon waveguide consisting of a similarly patterned central beam region surrounded by phononic bandgap shielding on either side. The phonon waveguide is designed to have large extrinsic coupling $\gamma_e$ to the nanobeam cavity mode, leading to hybridization of the localized cavity mode and the nearly-resonant phonon waveguide modes. The optical coupling waveguide is fabricated in the near-field of the nanobeam cavity (bottom beam of Fig.~\ref{fig1}b), allowing for evanescent coupling of laser light into and out of the cavity.  A single optical fiber taper is used to couple light into the on-chip coupling waveguide, and a photonic crystal mirror is etched in to the end of the optical coupling waveguide so that light coupled into the nanobeam cavity can be recollected by the optical fiber taper as per Ref.~\cite{Groeblacher2013}. 

Figure~\ref{fig1}c shows the band structure of the phonon waveguide with silicon slab thickness $d=220$~nm and lattice constant $a=a_0=480$~nm, numerically simulated using the finite-element method~\cite{comsol}. For clarity, only the bands with even $z-$symmetry (out-of-plane direction) are shown, with the green curve corresponding to the breathing mode waveguide band.  For a lattice constant of $a_0=480$~nm this band has a bandwidth of approximately $1$~GHz surrounding the $6$~GHz frequency of the localized acoustic mode of the nanobeam cavity.  Scaling of the planar waveguide dimensions by scale factor $s^{\prime}=a/a_0$, where $a$ is the new lattice constant, results in an approximate scaling of the waveguide bands by $(s^{\prime})^{-1}$. This is a good approximation for the lower lying bands of thin structures in which $d/a_0 \leq 1$.  As such, here we plot the bandstructure in terms of a (unitless) normalized frequency $s=(f/6$~GHz$)s^{\prime}$, where $f$ is the physical waveguide band frequency.  With the phonon waveguide terminated on one end by the nanobeam cavity and on the other end by cutting out a slot in the center beam, the continuous waveguide band is transformed into discrete waveguide resonances with free spectral range $f_{\rm FSR}=v_g/(2l)$, where $v_g=|d\omega/dk|$ is the group velocity of the band and $l$ is the length of the waveguide. For those waveguide modes spectrally located within a bandwidth $\gamma_e$ of the localized cavity mode, optical excitation of the optomechanical cavity can be used to detect phonons propagating in the waveguide section of the device. 

Optical excitation and detection of the waveguide phonon resonances are performed using a tunable narrowband laser with frequency ($\omega_p$) blue-detuned from the optical cavity frequency ($\omega_c$) by the mechanical cavity frequency ($\omega_m$), $\Delta \equiv \omega_p - \omega_c \approx \omega_m$.  Efficient Stokes scattering of the laser pump field occurs for the mechanical breathing mode resonances of the cavity-waveguide system which lie within an optical cavity linewidth ($\kappa$) of the bare mechanical frequency $\omega_m$.  For the blue-detuned sideband pumping employed throughout this work, the mechanical modes also experience amplification by the optical pump wave corresponding to a negative (phonon) decay rate of $\gamma_{\rm OM}=-4g_0^2n_c/\kappa$, where $g_0$ is the vacuum optomechanical coupling rate characterizing the strength of the radiation pressure interaction~\cite{cool1}, $n_c$ is intra-cavity photon number determined by the optical pump power, and $\kappa$ is optical cavity damping rate. 

All experiments are performed at room temperature and atmospheric pressure. The reflected optical signal from the cavity is detected on a photodetector, with the beating of the pump wave and resulting Stokes waves yielding a photocurrent spectrum containing the (mostly) thermal motion of the mechanical resonances of the cavity-waveguide system.  Figure~\ref{fig1}d shows a typical microwave photocurrent spectrum of a fabricated device with a waveguide length $l=145$~$\mu$m, cavity-to-waveguide coupling rate $\gamma_e/2\pi = 2.4$~MHz, and loaded optical cavity linewidth $\kappa/2\pi = 0.8$~GHz.  A series of hybridized modes with free-spectral range $f_{\rm FSR}=4$~MHz are observed around the main cavity resonance peak at $\omega_m/2\pi = 6.017$~GHz, from which one can infer a waveguide group velocity of $v_g=1160$~m/s. 
In order to trace out the group velocity evolution along the entire breathing mode waveguide band a set of devices is fabricated with the planar dimensions of the waveguide scaled by different scaling factors $s^{\prime}=a/a_0$.  The nanobeam cavity parameters are held fixed. In this way the waveguide band can be shifted in frequency relative to the nominally fixed cavity, allowing us to effectively scan across the whole waveguide band using the cavity as a frequency (and spatial mode) filter. Figure~\ref{fig1}e shows the measured group velocity evolution along the waveguide band as inferred from the measured $f_{\text{FSR}}$ of the waveguide resonances versus normalized frequency $s$.  The green curve corresponds to the theoretically calculated group velocity curve. We note that in the lower frequency portion of the band there exists two $k$-points for each frequency, resulting in two group velocity curves, a feature also seen in the measured data. In analogy to the slow light in photonic crystals~\cite{onchip_true_delay_line2}, here we measure slow phonon group velocities~\cite{wg4} near both band edges as well as at the band minimum occurring near $k \approx 0.5(\pi/a)$.  The minimum measured phonon group velocity is as small as $150$~m/s, a reduction by a factor of $40$ compared to the group velocity of transverse acoustic waves in bulk silicon at room temperature.  

\begin{figure*}[btp]
\begin{center}
\includegraphics[width=2\columnwidth]{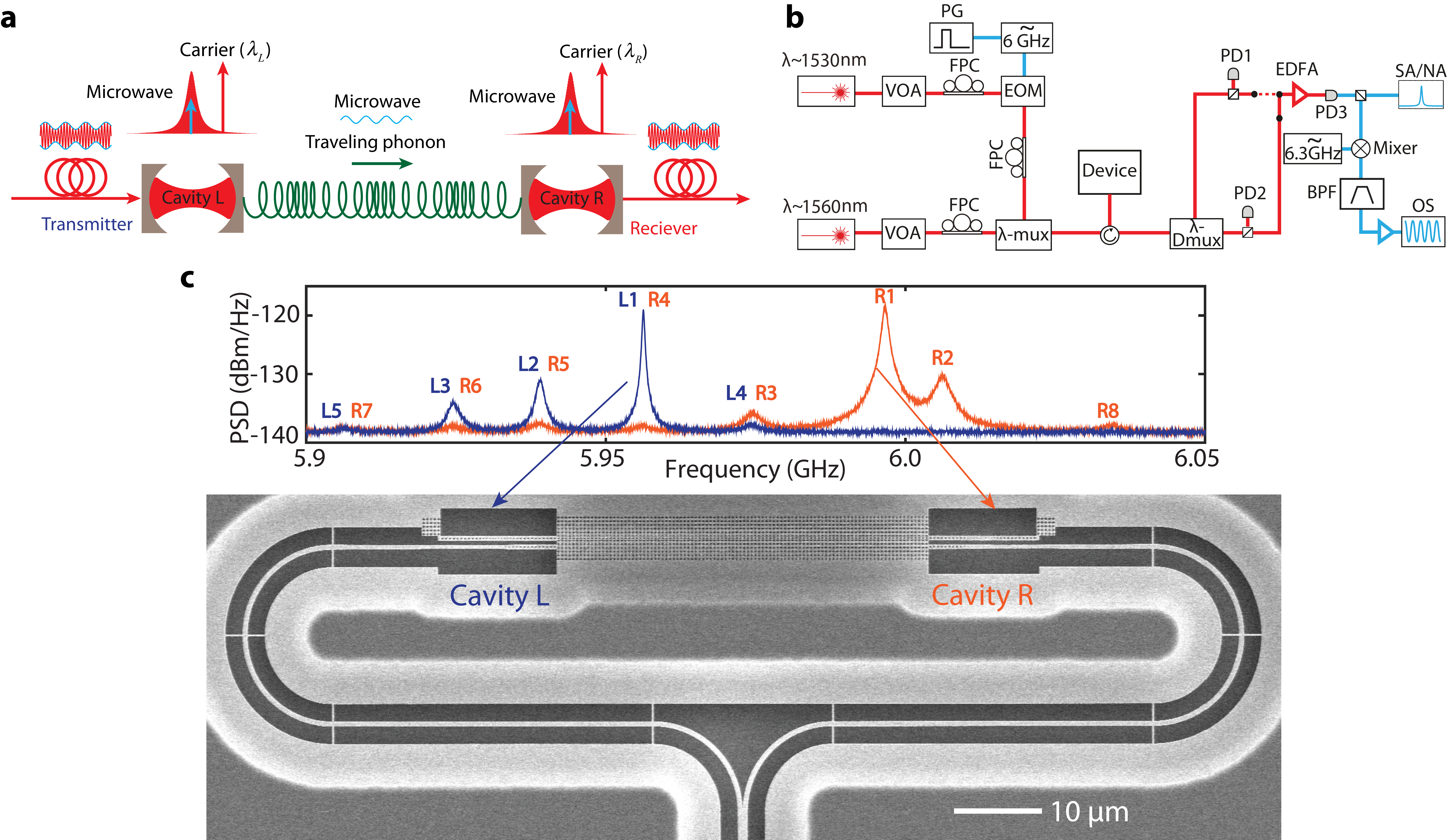}
\caption{ \textbf{Microwave signal processing using optomechanical cavity-waveguide system.} \textbf{a}, An optomechanical scheme for microwave signal processing relying on phonon routing. \textbf{b}, Schematic of the experimental set-up. VOA, variable optical attenuator; FPC, fiber polarizer controller; PG, pulse generator; EOM, elector-optic modulator; $\lambda$-(D)mux, wavelength (de-)multiplexer; PD, photodetector; EDFA, erbium doped fiber amplifier; SA, spectrum analyzer; NA, network analyzer; BPF, bandpass filter; OS, oscilloscope. \textbf{c}, SEM image of the double-cavity device, showing the 'Y'-branch optical splitter which feeds both left and right optical cavities and the intermediary mechanical waveguide that couples the left and right cavities together. The mechanical spectrum of the device as measured from cavity $L$ (blue curve) and $R$ (orange curve) is shown in the top inset.}
\label{fig2}
\end{center}
\end{figure*}

\section{Optomechanical Microwave Delay Line and Filter}
\label{sec:delay_line_filter}
We next study the utility of cavity-waveguide optomechanical systems for coherent microwave-over-optical signal processing.  Similar signal processing based on stimulated Brillouin scattering has recently been investigated in photonic nanoscale waveguides~\cite{rakich,sbs2,sbs3}.  Here we employ cavities as localized sites for enhancing the interaction of photons and phonons, and phonon waveguides for the routing and filtering of microwave signals.  The basic circuit, shown schematically in Fig.~\ref{fig2}a, consists of two optomechanical cavities, labeled $L$ and $R$, which are connected by a common phonon waveguide. The microwave signal encoded in the single sideband of an optical carrier with wavelength $\lambda_L$, is fed into cavity $L$ which coherently drives the localized cavity breathing mode through the radiation-pressure force. The excited phonons then couple into the phonon waveguide and propagate to cavity $R$, providing the desired delay and filter functions for the microwave signal. In cavity $R$, the arriving phonons modulate the internal optical field, re-encoding the microwave signal on an outgoing carrier at wavelength $\lambda_R$. 

A schematic of the experimental set-up used to measure the two-cavity system is shown in Fig.~\ref{fig2}b. SEM images of the cavity and waveguide elements as they are arranged in the optomechanical circuit are shown in Fig.~\ref{fig2}c. An adiabatic fiber-to-chip coupling waveguide followed by a 'Y'-branch splitter is used to couple laser light into and out of each of the cavity devices using a single optical fiber taper.  The two nanobeams are designed to have similar mechanical cavity frequencies but different optical resonant frequencies so that they may be simultaneously and independently probed using two separate laser sources (see App.~\ref{app:methods}).     

A typical mechanical spectrum of the double-cavity system with phonon waveguide length $l=43$~$\mu$m is shown in Fig.~\ref{fig2}c. For this device the two fundamental optical cavity modes $O_{L(R)}$ have wavelength $\lambda_{L(R)}=1539(1559)$~nm and linewidth $\kappa_{L(R)}=2\pi\times 1.15(0.80)$~GHz.  The blue curve corresponds to blue-detuned sideband pumping of cavity $L$ ($\Delta_{L}/2\pi = 6$~GHz; $P_{pL} = 180$~$\mu$W; $n_{cL} =1200$) and the laser resonant with cavity $R$ turned off.  The orange curve is for similar pumping of cavity $R$ ($\Delta_{R}/2\pi = 6$~GHz; $P_{pR} = 50$~$\mu$W; $n_{cR} = 360$).  The two spectra are seen to share a number of common resonances, a result of the connecting phonon waveguide.  We label the mechanical resonances as $L_j$ ($R_k$), with the index $j$ ($k$) indicating the relative optical coupling strength of the resonance to cavity $L$ ($R$).  The most intense mechanical resonance peaks in the cavity spectra, $L_1$ ($=R_4$) and $R_1$, are found to have vacuum optomechanical coupling rates of $g_{0L_1}/2\pi=0.85$~MHz to optical mode $O_{L}$ and $g_{0R_1}/2\pi=1.39$~MHz to optical mode $O_{R}$, respectively, determined from the variation of the mechanical linewidth with pump power.  The intrinsic $Q$-factor of the measured mechanical resonances are all approximately $1500$ as determined from the resonance peak linewidths at low optical probe power ($\gamma_{i}/2\pi \approx 3.7$~MHz).  


\begin{figure}[btp]
\begin{center}
\includegraphics[width=\linewidth]{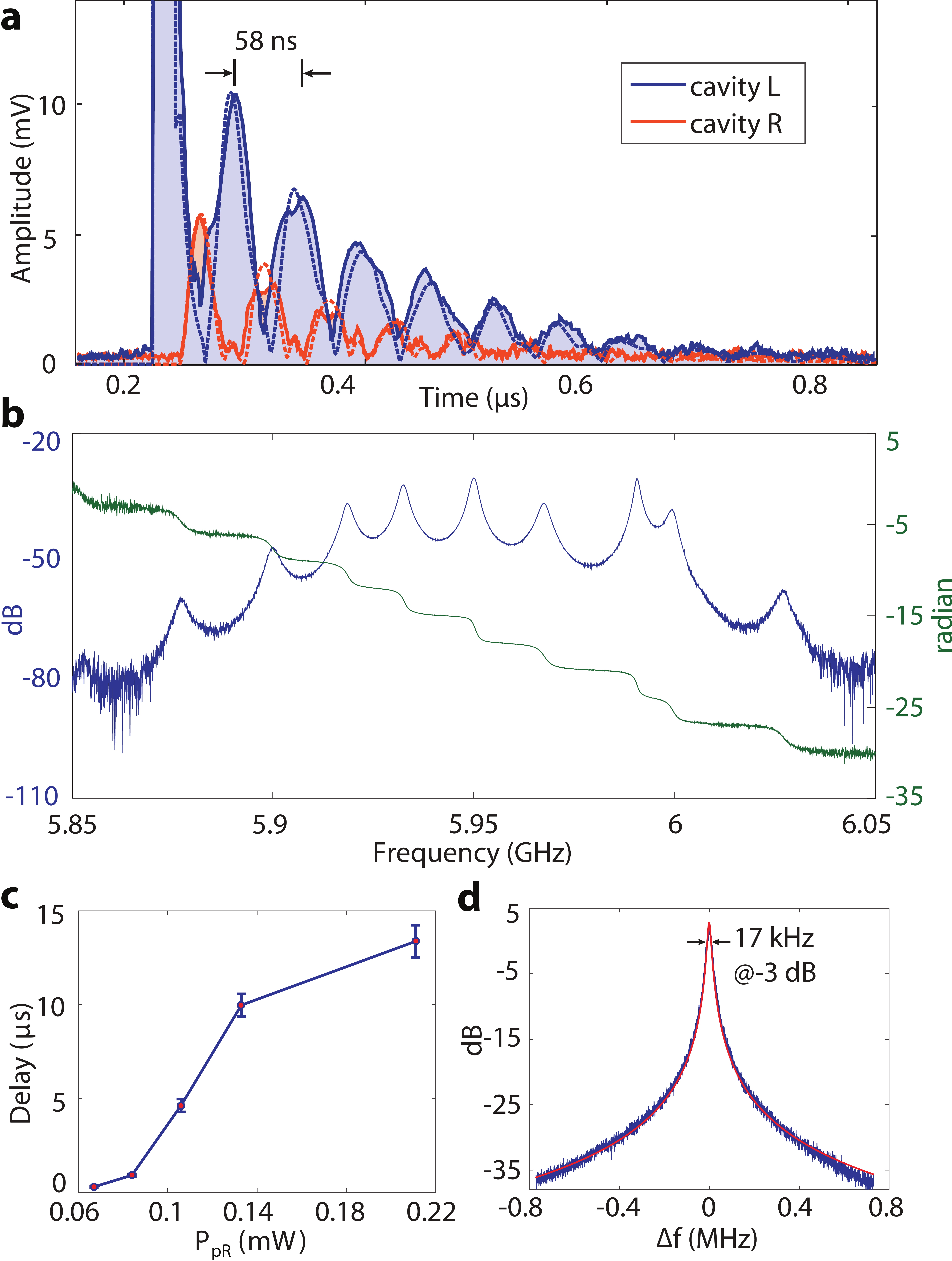}
\caption{\textbf{Pulsed and CW signal propagation.} \textbf{a}, A short phonon pulse excited by a $20$~ns microwave pulse of frequency $5.952$~GHz in cavity $L$ is detected in the two optomechanical cavities as it bounces between them. Dashed lines are simulated results. \textbf{b}, CW microwave propagation from cavity $R$ to cavity $L$ is characterized by $S$-matrix component $S_{RL}$ as measured by a network analyzer (see App.~\ref{app:methods}). Blue curve is $|S_{RL}|^2$ (in dB) and green curve is the phase of $S_{RL}$. \textbf{c}, Microwave delay at peak of $R_1$ resonance ($5.991$~GHz) versus pump power in cavity $R$ ($P_{pR}$) with fixed pump power in cavity $L$ ($P_{pL}=0.2$~mW) and microwave signal power $P_s=-20$~dBm. \textbf{d}, $|S_{RL}|^2$ spectrum around $R_1$ resonance corresponding to peak delay of $13.3$~$\mu$s. Red curve is a Lorentzian fit. }
\label{fig3}
\end{center}
\end{figure}

We next study the propagation of short phonon pulses within the double-cavity system. In this experiment the two cavities are simultaneously being probed with two separate continuous wave lasers ($P_{pL}=80$~$\mu$W, $n_{cL} = 530$; $P_{pR}=50$~$\mu$W, $n_{cR} = 360$), each tuned to the blue sideband of their respective cavity resonance.  A $20$~ns microwave pulse is used to phase modulate the laser light input to cavity $L$ at a frequency of $5.952$~GHz resonant with the $L_1$ mechanical resonance, exciting a coherent phonon population in cavity $L$ (see App.~\ref{app:methods}).  Figure~\ref{fig3}a shows the optically-transduced signal in both cavities after the initial excitation pulse.  For pulses with bandwidth larger than the free-spectral range (FSR) of the waveguide modes connecting the two cavities, the delay associated with phonons traversing the waveguide can be resolved and is given by, $\tau=1/2f_{\rm FSR}$.  Multiple bounces of the initial phonon pulse are seen in the signal from each cavity, with a period between bounces of $58$~ns (corresponding to $\tau=29$~ns) consistent with the measured FSR of $17$~MHz. The pulse propagation can be simulated with coupled mode equations (see App.~\ref{app:A}), and are shown as dashed lines in Fig.~\ref{fig3}a. From a fit of the model to the measured pulsed response, the extrinsic mechanical cavity-to-waveguide coupling rate is estimated to be $\gamma_{eL(R)}/2\pi=35(26)$~MHz for the $L$ ($R$) cavity, corresponding to an internal efficiency of phonon transmission from cavity $L$ to $R$ of $89\%$ ($67\%$) excluding (including) phonon waveguide decay. 


Continuous-wave (CW) measurements of the same device are presented in Fig.~\ref{fig3}b in which the frequency ($\omega$) of the phase modulation applied to the laser light sent into cavity $L$ is varied.  The displayed $S-$matrix component $S_{RL}$ (see App.~\ref{app:A}) is for an average optical pump power of $P_{pL(R)}=80$~$\mu$W for both cavities. Signal delay in this case can be inferred from the phase $\phi$ of $S_{RL}$, $\tau=-\frac{d\phi}{d\omega}$, where $\tau$ now includes both the delay associated with propagation through the phonon waveguide and the resonant storage time in each cavity.  Unlike previously studied optomechanically-induced transparency in a single cavity~\cite{eit1,eit2,eit3}, where delay is associated with reduced transmission because of the destructive interference between direct optical transmission and phonon mediated optical transmission, here the delay purely relies on phonons as no light directly propagates through the waveguide and one can simultaneously realize high transmission efficiency and large delay.  For blue-detuned sideband pumping, which results in parametric amplification of the phonon signal, the peak optical transmission gain between cavities for the phonon-waveguide mediated process can be greater than unity, and is given for excitation of mechanical resonance $L_j$ ($=R_k$) by,
 
\be\label{}
G_{j,k,\rm max}=\frac{4C_{L_j} C_{R_k}}{(1-C_{L_j}-C_{R_k})^2},
\ee

\noindent where $C_{L_j(R_k)}=|\gamma_{{\rm OM},L_j(R_k)}|/\gamma_i$ is the cooperativity between the $L_j$ ($=R_k$) mechanical resonance and the $O_{L(R)}$ optical cavity mode. This is the internal gain, and neglects optical loss outside of the cavities and photodetector inefficiency.  

Figure~\ref{fig3}c shows the measured delay for a $5.991$~GHz microwave signal resonant with the $R_1$ mode as a function of the optical pump power sent into cavity $R$ for fixed pump power $P_{pL}=0.2$~mW in cavity $L$.  The delay is tunable with optical pump power, reaching a value as large as $\tau=13.3$~$\mu$s for $P_{pR}=0.21$~mW.  Figure~\ref{fig3}d displays the measured $S-$matrix amplitude spectrum corresponding to the longest measured delay, showing a narrow $3$-dB bandwidth of $17$~kHz and a peak internal gain of $3$~dB.  Viewed as a tunable bandwidth microwave filter~\cite{filter_review,filter2,rakich}, this device realizes an out of band rejection of up to $70$~dB and a thermal-noise-limited sensitivity of $-30$ dBm referred to the microwave input of the electro-optic phase modulator (see App.~\ref{app:C}).   

\begin{figure}[btp]
\begin{center}
\includegraphics[width=\columnwidth]{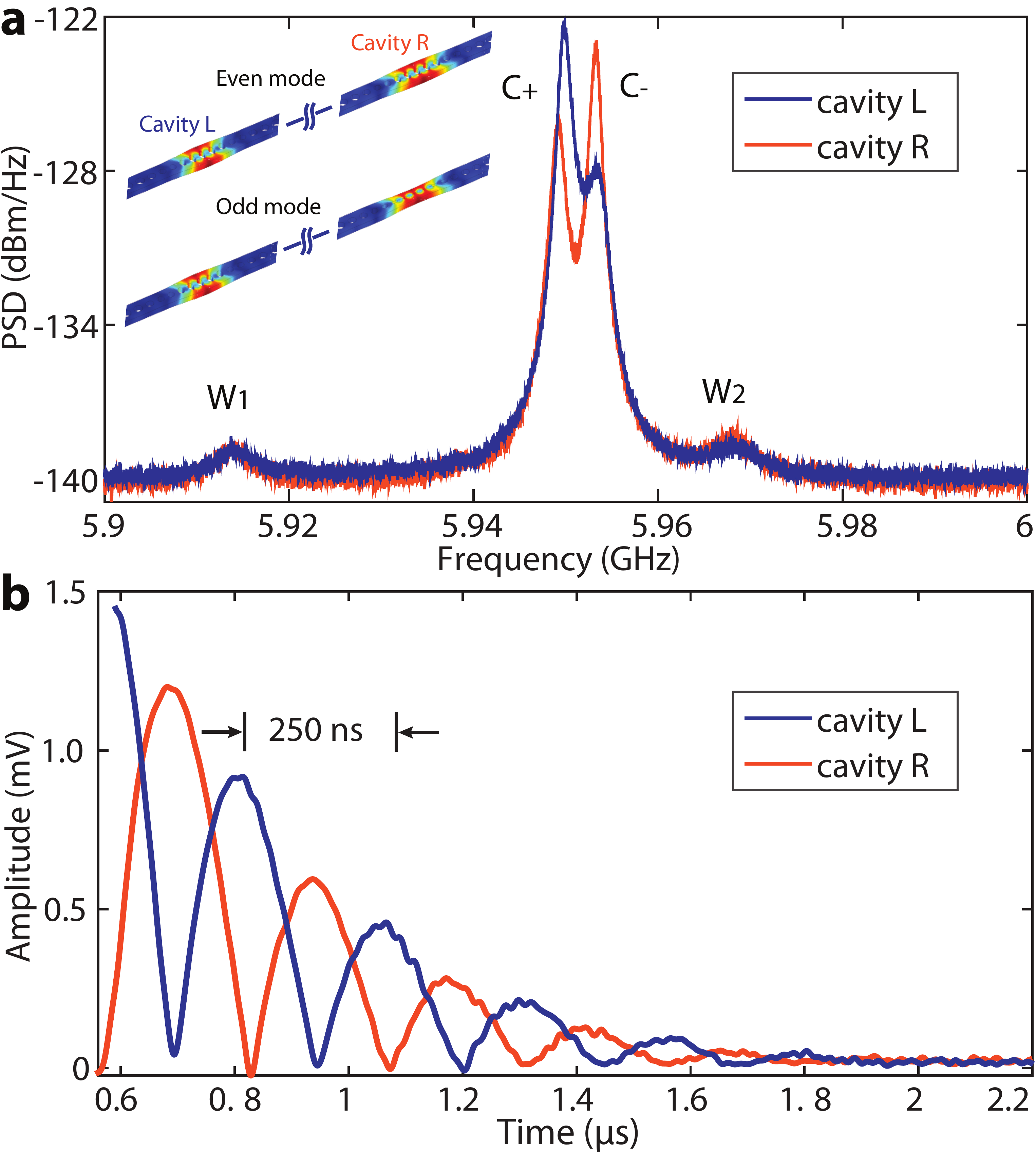}
\caption{ \textbf{Waveguide-mediated distant mechanical coupling.} \textbf{a} Mechanical spectrum of a device with two distantly-coupled cavities as measured from cavity $L$ (blue) and $R$ (orange). Modes $C_+$ and $C_-$ are the (approximately) even and odd hybridized mechanical cavity modes with the simulated modal profile shown in the inset. The weakly transduced $W_1$ and $W_2$ resonances are waveguide modes. \textbf{b} Measured phonon Rabi oscillation between the two distant cavities.}
\label{fig4}
\end{center}
\end{figure}

\section{Waveguide-Mediated Distant Mechanical Coupling}
\label{sec:cavity_cavity_coupling}
In addition to guiding the propagation of acoustic waves, phonon waveguides may also be used to mediate coupling between distant optomechanical cavities via virtual phonons, i.e. without energy distribution in the mediating waveguide modes.  Such a scenario can be used to realize highly efficient signal conversion between cavity elements, or in the case of mechanical sensing, may be employed to realize new sensing modalities in which read-out and sensor are separated spatially.  As recently demonstrated for optical photons in photonic crystals~\cite{noda}, when the cavity-to-waveguide coupling rate is substantially smaller than the FSR of the waveguide modes, $\gamma_{e}/2\pi \ll f_{\text{FSR}}$, the excitation of real waveguide phonons is suppressed and only second-order virtual transitions between cavity and waveguide modes contribute to the inter-cavity coupling.  For nearly degenerate bare cavity modes the waveguide mediated inter-cavity coupling strength scales as $\bar{\gamma_e} = \sqrt{\gamma_{eL}\gamma_{eR}}$, and for $\bar{\gamma_e}>\gamma_i+\gamma_{\rm OM}$ such a configuration can provide both high internal (i.e. phonon routing between cavities) and external (i.e. phonon-photon exchange in the cavity) efficiency for the microwave filter demonstrated in the previous section.  

Figure~\ref{fig4}a shows the measured (thermal) mechanical spectra of a device with two optomechanical cavities coupled together by a $30$~$\mu$m long phonon waveguide which is designed to have $\gamma_{e}/2/pi \ll f_{\text{FSR}}$.  This device is similar to that in Fig.~\ref{fig2}c with one key difference: the central holes of the waveguide have been removed, reducing the cavity-to-waveguide coupling.  An optically-transduced spectrum measured through both the left ($L$) and right ($R$) optical cavities is shown.  The two hybridized mechanical cavity modes, labeled $C_+$ and $C_-$, have a frequency splitting $\Delta_{+-}/2\pi=4$~MHz and are separated from the nearest waveguide mode by $\delta = 17$~MHz.  Both of the $C_+$ and $C_-$ mechanical resonances are measured to have large optomechanical couplings of over $700$~kHz with both optical cavity modes, indicating strong mixing of the $L$ and $R$ mechanical cavity modes.  The weakly transduced waveguide modes are measured to have a FSR of $f_{\rm FSR}=54$~MHz.   

From the matrix of optomechanical rates between the hybridized mechanical modes and the (uncoupled) optical cavity modes we determine a waveguide-mediated coupling of $V/2\pi=1.94 \pm 0.06$~MHz and a bare mechanical mode splitting of $\Delta_{LR}/2\pi = 0.98 \pm 0.48$~MHz (see App.~\ref{app:D}).  The coupling rate between nearly degenerate $L$ and $R$ mechanical cavity modes can be shown to be given to lowest order by $V = \bar{\gamma_e}/\sin(\pi\delta/f_{\text{FSR}})$, yielding an average cavity-to-waveguide coupling of $\bar{\gamma_e}/2\pi \approx 1.62$~MHz, consistent with the numerically-simulated design value for this device.  Despite the distant separation between the two cavities, phonons can nevertheless tunnel between them via virtual excitation of the waveguide modes. To show this, a $50$~ns pulse of a $5.95$~GHz microwave modulation signal is applied to the optical laser beam input to optical cavity $L$, locally exciting a coherent phonon population in the left cavity.  Following the pulse, Fig.~\ref{fig4}b shows the resulting oscillation of the phonon population between the two cavities.  The measured period of oscillation is $250$~ns, corresponding to a Rabi frequency of $4$~MHz commensurate with the hybridized mode splitting $\Delta_{+-}/2\pi$. This result should be contrasted with the previously demonstrated phonon pulse bouncing shown in Fig.~\ref{fig3}, where the period is set by the free spectral range of the waveguide modes.

\section{Discussion}
\label{sec:disc}
The devices presented in this work, involving pairs of optomechanical cavity elements connected together by a phononic waveguide, demonstrate the feasibility of creating small optomechanical crystal circuits for microwave-over-optical signal processing and novel sensing modalities.  Advantages of the optomechanical approach are myriad.  For instance, the use of radiation pressure as opposed to the piezo-electric effect to couple to microwave phonons allows one to work with a wider variety of materials, such as silicon in the devices studied here, and to realize high efficiency of signal conversion.  Purely electromechanical nanoscale devices~\cite{Mohammadi2006, Mohammadi2012}, by comparison, suffer from capacitance and impedance mismatch which limits efficiency.  Implementation of two-dimensional optomechanical crystals, as has been recently demonstrated in Ref.~\cite{2D}, can also be used to create multi-tap phonon waveguide structures capable of realizing diverse spectral filter shapes and dispersion control without the need for additional electrical or photonic hardware.    

Further scaling of optomechanical circuits, to arrays of a large number of cavity elements, is also interesting.  A straightforward one-dimensional extension of the currently studied structures to a chain of mechanically-coupled optomechanical crystal cavities, for instance, can be used to study mechanical collective dynamics~\cite{1dcollective}.  An even richer set of physical phenomena and device functionality, however, may be realized in optomechanical circuits in which both phononic and photonic connectivity of cavity elements are employed.  For example, building on the tight-binding mechanical coupling of cavities demonstrated in this work, an acoustic and optical isolator may be realized using a phase-correlated optical pumping scheme in which the cavities are also coupled together by a photonic waveguide~\cite{ophwg2,circulator2}.  In this scheme~\cite{ab}, analyzed in App.~\ref{app:E}, the interference between optical and mechanical hopping amplitudes between the cavities picks up the phase difference between the laser pump beams at each cavity site, giving rise to the required non-reciprocal behavior.  With the help of post-processing methods, such as scanning probe nano-oxidation lithography~\cite{afm} which has been successfully employed to tune photonic crystal cavities~\cite{imamoglu}, the task of aligning both mechanical and optical resonances of the optomechanical cavities could be fulfilled.  This would allow, at least in moderate lattice sizes, the study of topological phases and effective magnetic fields for photons and phonons~\cite{manybody1,manybody2,fang}.

\begin{acknowledgments}
The authors would like to thank Justin Cohen and Sean Meenehan for help with device fabrication and design.  This work was supported by the AFOSR Hybrid Nanophotonics MURI, DARPA ORCHID and MESO programs, the Institute for Quantum Information and Matter, an NSF Physics Frontiers Center with support of the Gordon and Betty Moore Foundation, and the Kavli Nanoscience Institute at Caltech.
\end{acknowledgments}

\bibliographystyle{ieeetr}

\appendix

\section{Methods}
\label{app:methods}
\subsection{Fabrication}
The devices are fabricated using a silicon-on-insulator wafer with silicon device layer thickness of $220$~nm and buried-oxide layer thickness of 2~$\mu$m. The device geometry is defined by electron beam lithography followed by inductively-coupled-plasma reactive ion etching to transfer the pattern through the 220-nm silicon device layer. The devices are then undercut using an HF:$\rm H_2$O solution to remove the buried oxide layer, and cleaned using a piranha etch.

\subsection{Experimental Set-Up}
Two lasers are used to probe the two nanobeam cavities individually, one of which can be encoded with microwave signals by an electro-optic phase modulator. The output of the two lasers are combined in a wavelength multiplexer and coupled into the device through a tapered optical fiber. The on-chip optical waveguide is split into two paths to transmit light into the two nanobeam cavities through side-coupled optical couplers. The reflected light from the cavities is re-collected by the optical fiber and passed through an optical circulator to separate it from the input light. A fiber-based wavelength de-multiplexer with $20$~nm channel spacing is then used to separate the two cavity signals.  Finally, the reflected light is amplified by an erbium-doped fiber amplifier and sent to a high speed photodetector ($12$~GHz bandwidth). The electrical signal from the photodetector is measured either with a spectrum analyzer, network analyzer, or high speed oscilloscope.\\

\subsection{Pulse and CW measurements}
For the phonon pulse measurement, a pulse generator is used to gate the microwave source which drives the electro-optic modulator with a repetition rate of $20$~kHz. The photodetected optical signal is mixed down to $300$~MHz with a second microwave source which is locked to the one driving the electro-optic modulator. The down-converted signal is filtered, amplified and averaged $10^4$ times on an oscilloscope with sampling rate of $10$~GSa/s.  For the CW measurements a network analyzer is employed, with the microwave output of the analyzer (port 1) connected to the electro-optic modulator used to phase modulate the light sent into cavity $L$, and the photodetected optical signal reflected from the cavity $R$ is connected to port 2 of the analyzer.  We define the $S$-matrix component $S_{RL}=\bar{S}_{21}$, where $\bar{S}_{21}$ has been normalized to remove the effects of optical insertion loss into the devices, optical pre-amplification, and photodetector gain.

\section{Simulation of phonon pulse propagation}
\label{app:A}

In this section, we show propagation and bouncing of phonon pulses in the cavity-waveguide system (Fig.~3a) can be well simulated by a group of coupled mode equations using input-output formalism. The dynamics captured by the coupled mode equations is a phonon pulse travelling in a waveguide terminated by two cavities with bare mechanical frequency $\omega_{mL,R}$ and waveguide coupling rate $\gamma_{eL,R}$. We approximate $\omega_{mL,R}$ to be the frequency of cavity-dominated modes $L_1$ and $R_1$ in the simulation. Since the response time of the optical cavity is much shorter than that of the mechanical cavity, we can exclude the dynamics of optical modes from these equations. Thus, the coupled mode equations can be written as follows,
\begin{widetext}
\bqa
\label{fitL} \frac{db_L}{dt}=-(i\omega_{mL}+\frac{\gamma+\gamma_{eL}}{2})b_L- i g_{0L}\alpha_{0L}\alpha_{+L}^\star e^{-i\omega_st}\Theta(t)\Theta(\tau-t)+\sqrt{\gamma_{eL}}b_{{\rm in},L}(t),\\
\label{fitR} \frac{db_R}{dt}=-(i\omega_{mR}+\frac{\gamma+\gamma_{eR}}{2})b_R+\sqrt{\gamma_{eR}}b_{{\rm in},R}(t),\\
\label{bin1}b_{{\rm in},L}(t)=e^{-\alpha l}\big (\sqrt{\gamma_{eR}}b_{R}(t-t_w)-b_{{\rm in},R}(t-t_w)\big ),\\
\label{bin2}b_{{\rm in},R}(t)=e^{-\alpha l}\big(\sqrt{\gamma_{eL}}b_{L}(t-t_w)-b_{{\rm in},L}(t-t_w)\big),
\eqa
\end{widetext}
where $\alpha_{0L}$ and $\alpha_{+L}$ are the amplitudes of optical pump and its red sideband in the left cavity, $\tau$ is the duration of excitation pulse, $\omega_s$ is the frequency of pulse, $\Theta(t)$ is the Heaviside step function, $\gamma$ is the effective decay rate of the excited mechanical mode, $\alpha\approx\gamma/v_{g}$ is the waveguide loss rate, and $t_w=1/(2f_{\rm FSR})-1/(\gamma_{eL}+\gamma)-1/(\gamma_{eR}+\gamma)$ is the single trip time the pulse spent in the waveguide. 

From the mechanical spectrum we find $\gamma=2\pi\times 2.1$ MHz for $L_1$ mode (the main coherently-driven mode) during the pulse measurement; and by fitting the pulse tails detected in each cavity we find $\gamma_{eL}=2\pi\times34.7$ MHz and $\gamma_{eR}=2\pi\times25.5$ MHz. Using these parameters, $|b_L|$ and $|b_R|$ can be numerically calculated from the coupled mode equations and the proportional voltage signals are shown in Fig.~3a. The simulated result captures the main features of the measured pulse data. In particular, the pulse splitting observed from cavity $R$ is due to the fact that the pulse frequency is not in resonant with cavity $R$ and thus experiences destructive interference inside this cavity. 

The phonon transfer efficiency from cavity $L$ to cavity $R$ is about $e^{-\gamma/(2f_{\rm FSR})}\approx 67\%$. The phonon transfer efficiency from cavity to waveguide for cavity $L$ and $R$ is $\gamma_{eL(R)}/(\gamma_{eL(R)}+\gamma)\approx 94\%(92\%)$.

We summarize the measured mechanical mode parameters of the device in Fig.~3 in Table \ref{table I}, where $g_0$ for the $L_j(R_k)$ modes is with respective to $O_{L(R)}$ optical cavity modes.

\begin{table}
\caption{Mechanical mode parameters}\label{table I}
\begin{center}
\begin{tabular}{|l|l|l|l|} \hline  & $g_{0}/2\pi$ (MHz) & $\gamma_i/2\pi$ (MHz) & $\gamma_e/2\pi$ (MHz) \\ \hline $L_1$ & 0.85 & 3.7 &  \\ \cline{1-3} $L_2$ & 0.75 & 3.7 & 34.7 \\ \cline{1-3} $L_3$ & 0.63 & 3.9 &  \\ \hline $R_1$ & 1.39 & 3.6 & 25.5 \\ \hline\end{tabular}
\end{center}
\end{table}

\section{Derivation of the microwave $S-$matrix for the optomechanical cavity-waveguide system}
\label{app:B}

Here, we derive the $S-$matrix for a microwave signal traversing the optomechanical cavity-waveguide system. We assume the mechanical amplitude is small such that only the first-order optical sideband needs to be considered. In the next section, we will verify the small mechanical amplitude assumption. 

The Hamiltonian of the system under continuous wave operation involves two optical cavity modes $a_{L,R}$ with frequency $\omega_{cL,R}$ parametrically coupled to a common mechanical mode $b$,
\begin{widetext}
\be\label{H}
\hat H=\sum_{k=L,R}\hbar\omega_{ck}\hat a_k^\dagger\hat a_k+\hbar \omega_m\hat b^\dagger\hat b+\sum_{k=L,R} \hbar g_{0k}(\hat b^\dagger+\hat b)\hat a_k^\dagger\hat a_k+\sum_{k=L,R} i\hbar\sqrt{\kappa_{ek}}\alpha_{pk}e^{-i\omega_{pk}t}(\hat a_k-\hat a_k^\dagger),
\ee
\end{widetext}
where the last term is the pumps of the two optical cavities. For simplicity we assume the pumping lasers are blue-detuned from the cavity resonances which is true for all of our experiments. Suppose the pumping laser for cavity $L$ is modulated at frequency $\omega$ by a microwave signal, then the operators of the system can be decomposed into carriers and sidebands,
\be\label{decomp}
\hat a_k=\alpha_{0k}e^{-i\omega_{pk}t}+\alpha_{+k}e^{-i(\omega_{pk}-\omega)t}, \quad \hat b=\beta_-e^{-i\omega t},
\ee
where we only keep the red sideband of the pumping lasers because of rotating wave approximation, given the sideband resolved condition $\omega_m\gg \kappa_k$ of our device. Suppose the pumps are strong enough such that the carrier operators can be treated as static variables, then the equations of motion of the system can be derived after substituting Eq.~\ref{decomp} into Eq.~\ref{H},
\begin{gather}
\label{opt}  i\omega\alpha_{+k}=(i\Delta_k-\frac{\kappa_k}{2})\alpha_{+k}-ig_{0k}\alpha_{0k}\beta_-^\star-\sqrt{\kappa_{ek}}\alpha_{\rm in,k},\\
\label{mech} -i\omega\beta_-=-(i\omega_m+\frac{\gamma_i}{2})\beta_--\sum_k ig_{0k}\alpha_{0k}\alpha_{+k}^\star,
\end{gather}
where $\Delta_k=\omega_{pk}-\omega_{ck}\approx\omega_m$. Solving Eq.~\ref{opt} and Eq.~\ref{mech} in the frequency range $|\omega-\omega_m|\ll\kappa_k$, we obtain
\be
\label{coherentphn}\beta_-=\frac{ig_{0L}\sqrt{\kappa_{eL}}\frac{2}{\kappa_L}\alpha_{0L}}{i(\omega_m-\omega)+\frac{\gamma_i}{2}-\sum_k\frac{2g_{0k}^2|\alpha_{0k}|^2}{\kappa_k}}\alpha_{{\rm in},L}^\star,
\ee
\bqa
\label{linear}\alpha_{{\rm out,}R}&=&-\sqrt{\kappa_{eR}}\alpha_{+R}\\\nonumber
&=&-\frac{4g_{0L}g_{0R}\sqrt{\kappa_{eL}\kappa_{eR}}/(\kappa_L\kappa_R)\alpha_{0L}^\star\alpha_{0R}}{i(\omega_m-\omega)+\frac{\gamma_i}{2}-\sum_k\frac{2g_{0k}^2|\alpha_{0k}|^2}{\kappa_k}}\alpha_{{\rm in,}L}.
\eqa
From Eq.~\ref{linear}, peak optical gain at $\omega=\omega_m$ is 
\be\label{gain}
G_{\rm max}=\frac{|\alpha_{{\rm out,}R}|^2}{|\alpha_{{\rm in,}L}|^2}=\frac{4C_LC_R}{(1-C_L-C_R)^2},
\ee
where $C_{L(R)}=|\gamma_{{\rm OM},L(R)}|/\gamma_i$ is the coorperativity of mechanical mode $b$ with optical modes $a_{L(R)}$. 

Using the result of Eq.~\ref{linear}, the microwave signal transfer $S-$matrix can be derived
\bqa\label{smatrix}
&&S_{RL}\\\nonumber
&\equiv& \frac{V_{\rm NA, in}}{V_{\rm NA, out}}\\\nonumber
&=&\frac{\eta_{oL}\eta_{oR}G_eG_{\rm EDFA}(i\hbar\omega_{cR}\omega_m/\sqrt{\kappa_{eR}})\alpha_{{\rm out,}R}\alpha_{0R}^\star}{(2V_\pi/\pi)\big(\alpha_{{\rm in,}L}/(i\omega_m\alpha_{0L}/\sqrt{\kappa_{eL}})\big)}\\\nonumber
&=&\frac{\eta_{oL}\eta_{oR}G_eG_{\rm EDFA}}{2V_\pi/\pi}\frac{4g_{0L}g_{0R}\hbar\omega_{cR}\omega_m^2/(\kappa_L\kappa_R)|\alpha_{0L}|^2|\alpha_{0R}|^2}{i(\omega_m-\omega)+\frac{\gamma_i}{2}-\sum_k\frac{2g_{0k}^2|\alpha_{0k}|^2}{\kappa_k}}
\eqa   
where $V_{\rm NA, out}$ and $V_{\rm NA, in}$ are the output and detected electrical voltage of the network analyzer respectively, $\eta_{oL,R}$ is the optical loss of the input and output ports of the device and fiber respectively, $G_{\rm EDFA}$ and $G_e$ are the gain coefficients of EDFA and photodetector respectively, and $V_\pi$ is the voltage required to produce a phase shift of $\pi$ of the electro-optic modulator.

\section{Analysis of linear operation and noise characteristics of the optomechanical microwave filter/delay line}
\label{app:C}

In this section, we examine the assumption of weak mechanical amplitude under strong optical pump and analyze the optomechanical microwave filter/delay line performance in terms of linearity and noise characteristics. We find that thermo-optic effect constrains the mechanical amplitude due to saturation of the optomechanical gain.  This effect sets the linear operation range and the suppression of the mechanical thermal noise.

The thermo-optic effect induced optical cavity frequency shift can be described by the following equations~\cite{thermopt}
\bqa
\label{therm1}\delta\omega_c=-\omega_cn_{\rm Si}(T_0)\frac{dn_{\rm Si}(T_0)}{dT}A\delta T,\\
\label{therm2}\delta T=\frac{r\varsigma c^2}{n_{\rm Si}(T_0)^2V_{\rm TPA}}n_c^2,
\eqa
where $n_c$ is cavity photon number, $n_{\rm Si}$, $r$, and $\varsigma$ is the refractive index,  thermal resistance, and two-photon absorption coefficient of silicon respectively, $c$ is the speed of light, $V_{\rm TPA}$ is the cavity volume for two-photon absorption, and $A$ is a perturbation theory coefficient $A=\frac{\int_{\rm Si}|\bold E(\bold r)|^2d\bold r}{\int n_{\rm Si}(T_0)^2|\bold E(\bold r)|^2d\bold r}$. Substituting Eq.~\ref{therm2} into Eq.~\ref{therm1}, and using the parameters of silicon given in Ref.~\cite{thermopt}, along with $A\approx7.5\times10^{-2}$, $V_{\rm TPA}\approx(\lambda/n_{\rm Si}(T_0))^3$, we have
\be\label{fshift}
\delta\omega_c=\xi n_c^2, \quad \xi\approx -33.9~\rm Hz.
\ee

We proceed to include the term of thermo-optic effect (Eq.~\ref{fshift}) to the equations of motion (Eqs.~\ref{opt} and~\ref{mech}), which are then modified to be
\begin{widetext}
\bqa
\label{opt2}  i\omega\bar\alpha_{+}=\Big(i\big(\Delta+\xi(|\alpha_0|^2+|\bar\alpha_+|^2)^2\big)-\frac{\kappa}{2}\Big)\bar\alpha_{+}-ig_0\alpha_{0}\bar\beta_-^\star,\\
\label{mech2} -i\omega\bar\beta_-=-(i\omega_m+\frac{\gamma_i}{2})\bar\beta_-- ig_0\alpha_{0}\bar\alpha_{+}^\star-\sqrt{\gamma_i}\beta_{\rm in}, \quad \beta_{\rm in}=\sqrt{\gamma_i n_{\rm th}}/2,
\eqa
\end{widetext}
where we have denoted $\bar\alpha_+$ and $\bar\beta_-$ as the static value of the  corresponding operators without input optical sideband signal and we have included the mechanical thermal noise input. Also, we specifically consider the operation with the $R_1$ mode, and thus ignore optical cavity $L$ which has much weaker coupling with $R_1$ compared to optical cavity $R$. Eqs.~\ref{opt2} and~\ref{mech2} can only be solved numerically for a generic pump condition. To reveal the thermo-optic effect on the mechanical amplitude, we consider a special pump condition corresponding to the original threshold of mechanical self-oscillation, i.e. $4g_0^2|\alpha_0|^2/(\kappa\gamma_i)=1$. In this case, we can analytically solve for the down-converted photon number and mechanical amplitude at $\omega=\omega_m$ from Eqs.~\ref{opt2} and~\ref{mech2}, assuming $\Delta=\omega_m$ and $|\bar\alpha_+|\gg|\alpha_0|$,
\bqa
|\bar\alpha_+|^2=(\frac{\kappa\gamma_in_{\rm th}}{4\xi^2})^{1/5},\\
|\bar\beta_-|^2=\frac{\kappa}{\gamma_i}(\frac{\kappa\gamma_in_{\rm th}}{4\xi^2})^{1/5}+n_{\rm th}.
\eqa
For $\kappa=2\pi\times0.8$ GHz, $\gamma_i=2\pi\times3.6$ MHz, $g_0=2\pi\times1.39$ MHz, $n_{\rm th}=\frac{k_BT_0}{\hbar \omega_m}\approx1000$, we have $|\bar\alpha_+|^2\approx1900$, $|\alpha_0|^2\approx380$, and $|\bar\beta_-|^2\approx3.7\times10^5$. It is the optical resonance shift induced by thermo-optic effect that saturates optomechanical gain and prevents runaway of the mechanical amplitude at the threshold. 

Now we can estimate whether the mechanical amplitude is large enough to induce nonlinearity through excitation of higher order optical sidebands. The nonlinearity arises due to pump saturation and occurs when the amplitude of the first order optical sideband significantly deviates from being linearly proportional to the mechanical amplitude, i.e. approximation $J_1(z)\approx\frac{z}{2}$ breaks down~\cite{largemech}, where  $z=g_0\sqrt{4|\bar\beta_-|^2+2}/\omega_m$ is the normalized mechanical amplitude. For $|\bar\beta_-|^2\approx3.7\times10^5$ calculated above, the deviation is only about 1$\%$. In the experiment, we find for the largest pump power $P_{pL,R}\approx0.2$ mW, $|\bar\beta_-|^2\approx2.0\times 10^5$, which gives $z=0.18$ and a linear deviation of $0.4\%$. As a result, scattering into higher order optical sidebands does not need to be included in Eq. (\ref{opt2}) and (\ref{mech2}). 

Next, we consider the response of the mechanical oscillator to a small input optical sideband signal by perturbative expansion of Eqs.~\ref{opt2} and~\ref{mech2}. In this case the coherent mechanical amplitude of Eq.~\ref{coherentphn} is modified to be 
\be
\label{coherentphnwthermo}\beta_-=\frac{ig_{0L}\sqrt{\kappa_{eL}}\frac{2}{\kappa_L}\alpha_{0L}}{i(\omega_m-\omega)+\frac{\gamma_i}{2}-\sum_k\frac{2g_{0k}^2|\alpha_{0k}|^2}{-2i\delta_k+\kappa_k}}\alpha_{{\rm in},L}^\star,
\ee
where $\delta_k=\xi(|\alpha_{0k}|^2+|\bar\alpha_{+k}|^2)^2$ is the thermo-optic-effect induced optical frequency shift. Eqs.~\ref{linear} and~\ref{smatrix} can be modified correspondingly. According to Eq.~\ref{coherentphnwthermo}, the effective mechanical loss rate is $\gamma_{\rm eff}=\gamma_i-\sum_k\frac{4g_{0k}^2|\alpha_{0k}|^2}{\kappa_k}+\sum_k\frac{4g_{0k}^2|\alpha_{0k}|^2}{\kappa_k}(\frac{\delta_k}{\kappa_k/2})^2$. The deviation from a linear response can be caused by the additional cavity photons from the input signal, and is characterized by the ratio $r=|\alpha_{+k}|^2/|\bar\alpha_{+k}|^2$ (in our device the contribution is mainly from $\alpha_{+R}$). At the theoretical self-oscillation threshold, we find the $1$~dB compression point of the $S-$matrix to be equal to a microwave power of -19 dBm (assuming $|\alpha_{0L}|^2=|\alpha_{0R}|^2$). In the experiment, we find for the largest optical pump power $P_{pL,R}=0.2$ mW (which is slightly above the self-oscillation threshold), the $1$~dB compression point occurs at a microwave signal power of -15 dBm. For reversed operation (cavity $R$ as input), the 1 dB compression point is reduced by a factor of  $\gamma_{{\rm OM},L}/\gamma_{{\rm OM},R}$ (assuming $|\alpha_{0L}|^2=|\alpha_{0R}|^2$).

\begin{figure}[btp]
\begin{center}
\includegraphics[width=0.8\columnwidth]{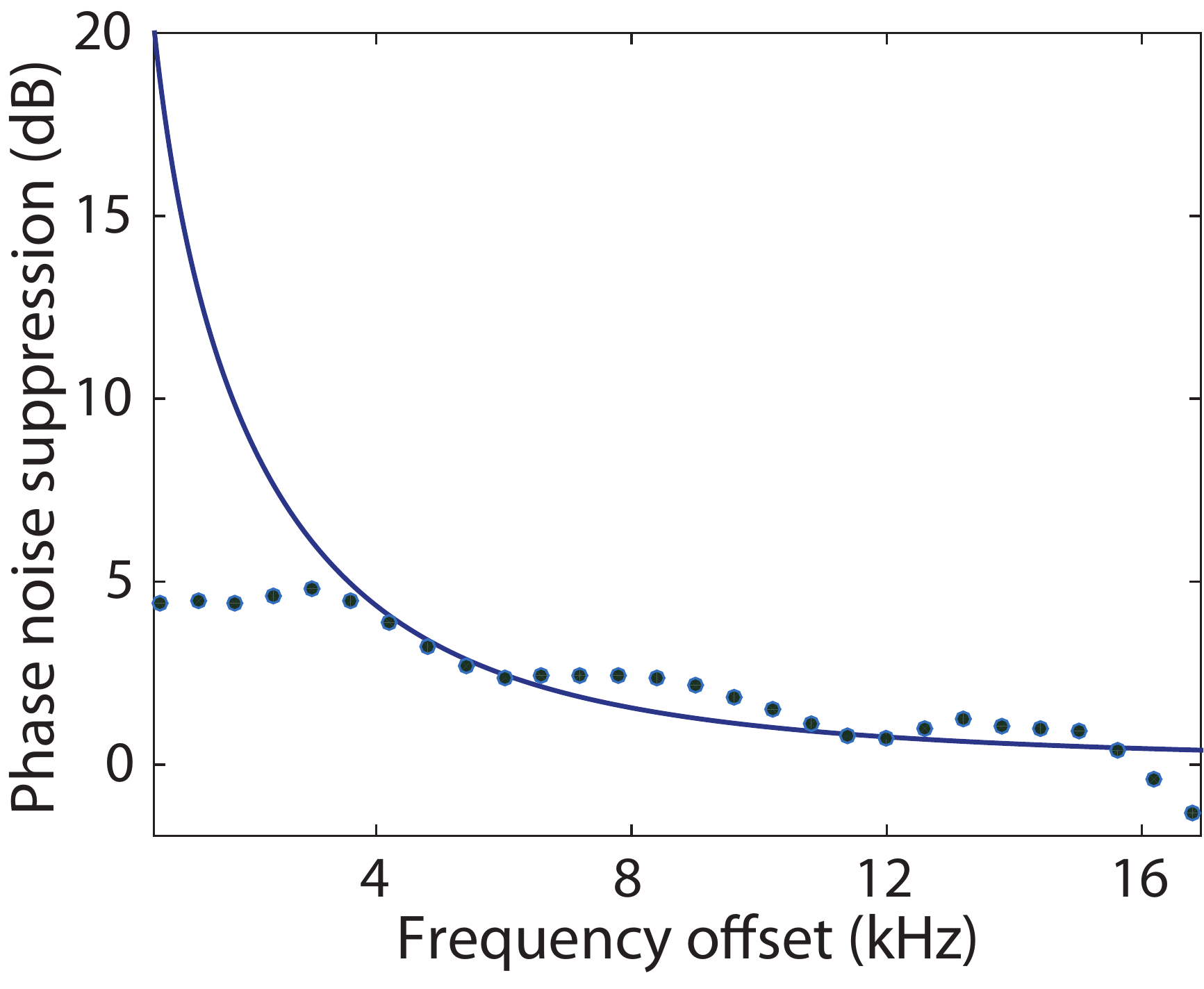}
\caption{Phase noise suppression ratio between microwave signal power of -15 dBm and -30 dBm at largest optical pump level. Solid line is theoretical value calculated using  Eq.~\ref{phasenoisecompression} and dots are experimental data.}
\label{suppfig1}
\end{center}
\end{figure}

We now analyze the noise characteristics of the optomechanical cavity-waveguide system.  The dominant form of noise is from thermally excited phonons in the system.  From the thermally-added mechanical noise refered to the input signal  $(\kappa_{e}/\kappa)^{-1}\gamma_in_{\rm th}/|\gamma_{{\rm OM}}|$~\cite{lambda1}, we define the noise-equivalent optical signal power 
\be
P_{\rm NE}=\frac{\kappa}{\kappa_e}\hbar\omega_c\frac{\gamma_i n_b}{\gamma_{\rm OM}}2\pi B,
\ee
where $B$ is the bandwidth of the coherent signal and all the quantities are referred to the input cavity. Then the noise-equivalent microwave signal power is
\be
V_{NE}^2=\frac{4}{\pi^2}\frac{P_{\rm NE}}{P_p}V_\pi^2.
\ee
We find that for the largest pump power when operating at $R_1$ resonance, if the input port is cavity $L$, the noise equivalent microwave power is -30 dBm; if the input port is cavity $R$, the noise equivalent microwave power reduces to -70 dBm because of the significantly enhanced $\gamma_{\rm OM}$ of cavity $R$ with $R_1$ mode.

For a self-oscillating mechanical oscillator, the intrinsic oscillator noise can be suppressed by the injection of an external coherent signal~\cite{vahala}. The suppressed phase noise (ignoring input signal noise) can be modeled by~\cite{phasenoise}
\be\label{phasenoisecompression}
\tilde S_{\phi\phi}(\omega-\omega_m)=\frac{1}{1+(\frac{\gamma_{\rm eff}}{\omega-\omega_m})^2\rho^2}S_{\phi\phi}(\omega-\omega_m),
\ee
where $S_{\phi\phi}(\omega)$ is the intrinsic phase noise spectral density without injection and $\rho=|\beta_-|/|\bar\beta_-|$ is the ratio between injected mechanical amplitude and free-running amplitude. Experimentally, we infer the phase noise from the measured noise power spectral density using the definition $S_{\phi\phi}(\omega)=S_{bb}(\omega)-\int S_{bb}(\omega)d\omega$. At the largest optical pump level, the phase noise suppression ratio between the microwave signal power -15 dBm (1 dB $S-$matrix compression point) and -30 dBm (noise equivalent power) is shown in Fig.~\ref{suppfig1}. The model of Eq.~\ref{phasenoisecompression} explains well the measured noise suppression level in the offset frequency ($\omega-\omega_m$) range between $\gamma_{\rm eff}/4$ and $3\gamma_{\rm eff}/4$ ($\gamma_{\rm eff}=2\pi\times 17$ kHz).

\section{Waveguide-mediated cavity-cavity coupling}
\label{app:D}
\subsection{Analytical derivation}
\label{app:D1}

When $\gamma_{eL,R}/2\pi\ll f_{\rm FSR}$, and all the waveguide modes are outside of the line width of the cavities, two mechanical cavities can acquire a waveguide mediated coupling. The generic Hamiltonian describing this case is 

\begin{widetext}
\be\label{Hw}
\hat H=\hbar\omega_{mL}\hat b_L^\dagger\hat b_L+\hbar\omega_{mR}\hat b_R^\dagger\hat b_R+\sum_{k} \hbar \omega_{k}\hat b_k^\dagger\hat b_k+\sum_{k} \hbar (g_{Lk}\hat b_L^\dagger\hat b_k+g_{Lk}^\star\hat b_L\hat b_k^\dagger)+\sum_{k} \hbar (g_{Rk}\hat b_R^\dagger\hat b_k+g_{Rk}^\star\hat b_R\hat b_k^\dagger),
\ee
\end{widetext}
where $b_k^\dagger(b_k)$ is the creation(annihilation) operator of the $k-$th waveguide mode,  $g_{Lk}=\sqrt{2\gamma_{eL}f_{\rm FSR}}$ and $g_{Rk}=(-)^k\sqrt{2\gamma_{eR}f_{\rm FSR}}$ are the coupling coefficients of the left and right mechanical cacity modes with the $k-$th waveguide mode, and the summation is over all waveguide modes. Note the relative sign between $g_{Lk}$ and $g_{Rk}$ comes from the symmetry of waveguide modes with respect to the center of the two cavities. 

In the degenerate cavity case, i.e. $\omega_{mL}=\omega_{mR}$, we can calculate the coupling $V$ between the two cavities mediated by the waveguide modes using second order perturbation theory as follows,
\bqa
\label{exactV}\frac{V}{2\pi}&=&\sum_{k=-\infty}^{\infty} \frac{\frac{g_{Lk}}{2\pi}\frac{g_{Rk}}{2\pi}}{\delta-kf_{\rm FSR}}\\\nonumber
&=&\frac{\sqrt{\gamma_{eL}\gamma_{eR}}}{2\pi} \sum_{k=-\infty}^{\infty} \frac{(-)^k}{\frac{\delta}{f_{\rm FSR}}\pi-k\pi}\\\nonumber
&=&\frac{\sqrt{\gamma_{eL}\gamma_{eR}}}{2\pi} \frac{1}{{\rm sin}\frac{\delta}{f_{\rm FSR}}\pi},
\eqa
where $\delta$ is the frequency difference between the degenerate cavity modes and their nearest waveguide mode. For the perturbation theory to be valid, we require 
\be\label{perturbcondition}
|g_{L,Rk}|/2\pi=\sqrt{2\gamma_{eL,R}f_{\rm FSR}}/2\pi\ll \delta.
\ee

We consider two special cases. First, when $\delta=f_{\rm FSR}/2$, i.e. the two cavity modes are in the center of two nearest waveguide modes, then $V=\sqrt{\gamma_{eL}\gamma_{eR}}$. In the other limit when $\delta\ll f_{\rm FSR}$, then $V\approx\sqrt{\gamma_{eL}\gamma_{eR}}\frac{f_{\rm FSR}}{\pi\delta}$. One can prove in this case, $V\ll \sqrt{2\gamma_{eL,R}f_{\rm FSR}}=|g_{L,Rk}|$, by taking into account of the condition of Eq.~\ref{perturbcondition}. Different from the previous case, here the contribution to the coupling can be almost exclusively attributed the nearest waveguide mode ($k=0$).

\subsection{Measurement modeling}
\label{app:D2}

\begin{figure}
\begin{center}
\includegraphics[width=0.8\columnwidth]{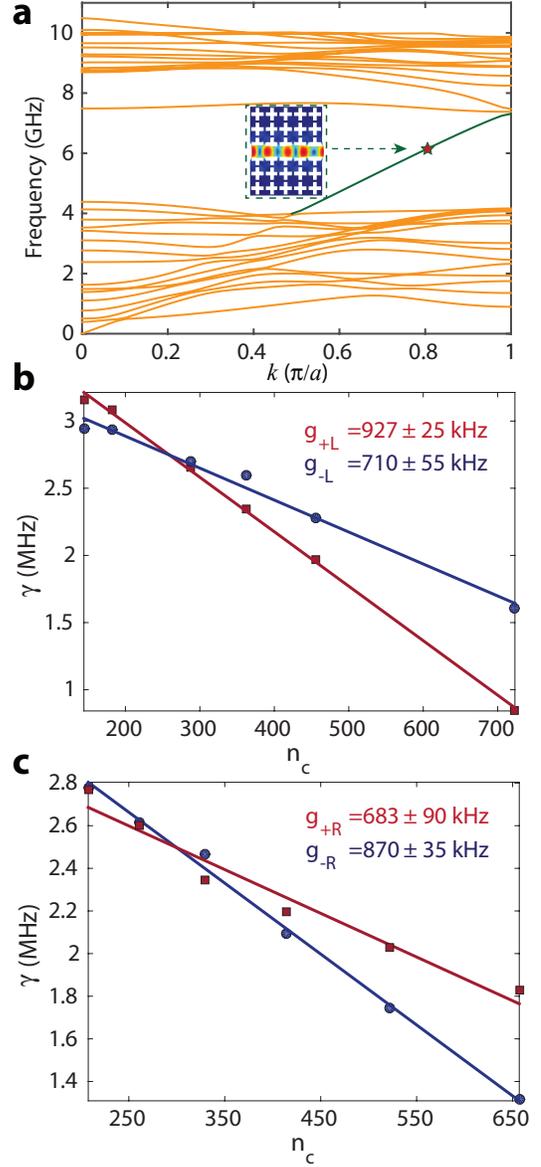}
\caption{\textbf{a} Band structure of shielded phonon waveguide without air holes. Inset is the modal profile with $k=0.8\pi/a$. \textbf{b,c} Characterization of optomechanical coupling between hybridized mechanical modes $C_+$ and $C_-$ with the optical modes of cavity $L$ (\textbf{b}) and $R$ (\textbf{c}). $\gamma=\gamma_i+\gamma_{\rm OM}$ is effective mechanical decay rate and $n_c$ is the cavity photon number.}
\label{suppfig2}
\end{center}
\end{figure}

We model how to estimate the phonon waveguide mediated coupling between two optomechanical cavities. In the device for demonstrating waveguide-mediated cavity coupling, we used a phonon waveguide without air holes. The band structure of the waveguide is shown in Fig.~\ref{suppfig2}a. Such a waveguide provides large phonon group velocity and thus large free spectral range in order to isolate cavity modes from waveguide modes. 

In the sample device shown in Fig.~4 of the main text, virtual phonons in waveguide mixes the mechanical cavity modes $M_L$ and $M_R$ of cavity $L$ and $R$ into hybridized modes $C_+$ and $C_-$. The coupling strength can be inferred by measuring the optomechanical coupling of hybridized modes $C_+$ and $C_-$ with the two optical modes $O_L$ and $O_R$ based on the following model. $C_+$ and $C_-$ can be expressed as linear superposition of $M_L$ and $M_R$, assuming ignorable energy distribution in the waveguide (as $C_{+,-}$ are well separated from waveguide modes), 
\bqa
C_+=\alpha_1M_L+\alpha_2M_R,\\
C_-=\beta_1M_L+\beta_2M_R.
\eqa
The superposition coefficients $\alpha_i$ and $\beta_i$ satisfy the following relation
\be\label{ratio}
|\frac{\beta_1}{\alpha_1}|=|\frac{\alpha_2}{\beta_2}|=\frac{2|V|}{\Delta_{LR}+\Delta_{+-}},
\ee  
where $V$ is the waveguide mediated coupling between $M_L$ and $M_R$, $\Delta_{LR}$ is the fabrication induced frequency difference between $M_L$ and $M_R$, and $\Delta_{+-}=\sqrt{\Delta_{LR}^2+4V^2}$ is the frequency difference between the hybridized modes $C_+$ and $C_-$. On the other hand, the ratio of the superposition coefficients is directly related to the measurable optomechanical coupling
\be
\frac{g_{+L}}{g_{-L}}=|\frac{\alpha_1}{\beta_1}|,\quad \frac{g_{+R}}{g_{-R}}=|\frac{\alpha_2}{\beta_2}|,
\ee
where, for example, $g_{+L}$ stands for the coupling between mechanical mode $C_+$ and optical mode $O_L$. 

We inferred the optomechanical couplings by measuring the pump dependent effective mechanical damping rate of $C_{+,-}$ (Fig.~\ref{suppfig2}b,c) using the relation $\gamma=\gamma_i-g^2n_c\frac{\kappa}{(\Delta-\omega_m)^2+\kappa^2/4}$, and obtained
\bqa
g_{+L}=927\pm25~{\rm kHz},\quad g_{-L}=710\pm55~\rm kHz\\
g_{+R}=683\pm90~{\rm kHz},\quad g_{-R}=870\pm35~\rm kHz
\eqa
As a result, the ratio of superposition coefficients is
\be\label{fit}
|\frac{\alpha_1}{\beta_1}|=1.30\pm0.14, \quad |\frac{\alpha_2}{\beta_2}|=0.79\pm0.14.
\ee
Along with $\Delta_{+-}/2\pi=4.0$ MHz as read from the spectrum, we find, according to Eq.~\ref{ratio},
\be
|V|/2\pi=1.94\pm0.06\ {\rm MHz, }\ \Delta_{LR}/2\pi=0.98\pm0.48\ {\rm MHz}.
\ee

To compare with the analytical formula of $V$ (Eq.~\ref{exactV}), in this device, we have $f_{\rm FSR}=54$ MHz and $\delta=17$ MHz. And using the measured $V/2\pi=1.94$ MHz, we find $\sqrt{\gamma_{eL}\gamma_{eR}}/2\pi=1.62$ MHz.

\section{Optical non-reciprocity based on distantly-coupled optomechanical cavities}
\label{app:E}

We provide theoretical analysis of achieving optical non-reciprocity in the distantly-coupled optomechanical cavities, and show its viability based on the typical parameters of our fabricated devices. In this case, the waveguide connecting the two cavities should support both guided mechanical and optical modes, and the two optomechanical cavities are designed to be identical. As already been demonstrated in our experiment and in Ref.~\cite{noda} respectively, such a waveguide can mediate a tight-binding type coupling for the mechanical and optical cavity modes. The Hamiltonian of the system can thus be written as follows,
\begin{widetext}
\bqa\label{H2}
\hat H&=&\sum_{k=L,R}\hbar\omega_{ck}\hat a_k^\dagger\hat a_k+J(a_L^\dagger a_R+a_La_R^\dagger)+\sum_{k=L,R}\hbar \omega_{mk}\hat b_k^\dagger\hat b_k+V(b_L^\dagger b_R+b_Lb_R^\dagger)\\\nonumber
&&+\sum_{k=L,R} \hbar g_{k}(\hat b_k^\dagger+\hat b_k)\hat a_k^\dagger\hat a_k+\sum_{k=L,R} i\hbar\sqrt{\kappa_{ek}}\alpha_{pk}e^{-i\omega_{p}t-i\varphi_k}(\hat a_k-\hat a_k^\dagger),
\eqa
\end{widetext}
where $J$ and $V$ are the waveguide mediated optical and mechanical coupling strength, and the last two terms are the optical pumps in the two cavities which have a same frequency and correlated phases.

\begin{figure*}
\begin{center}
\includegraphics[width=2\columnwidth]{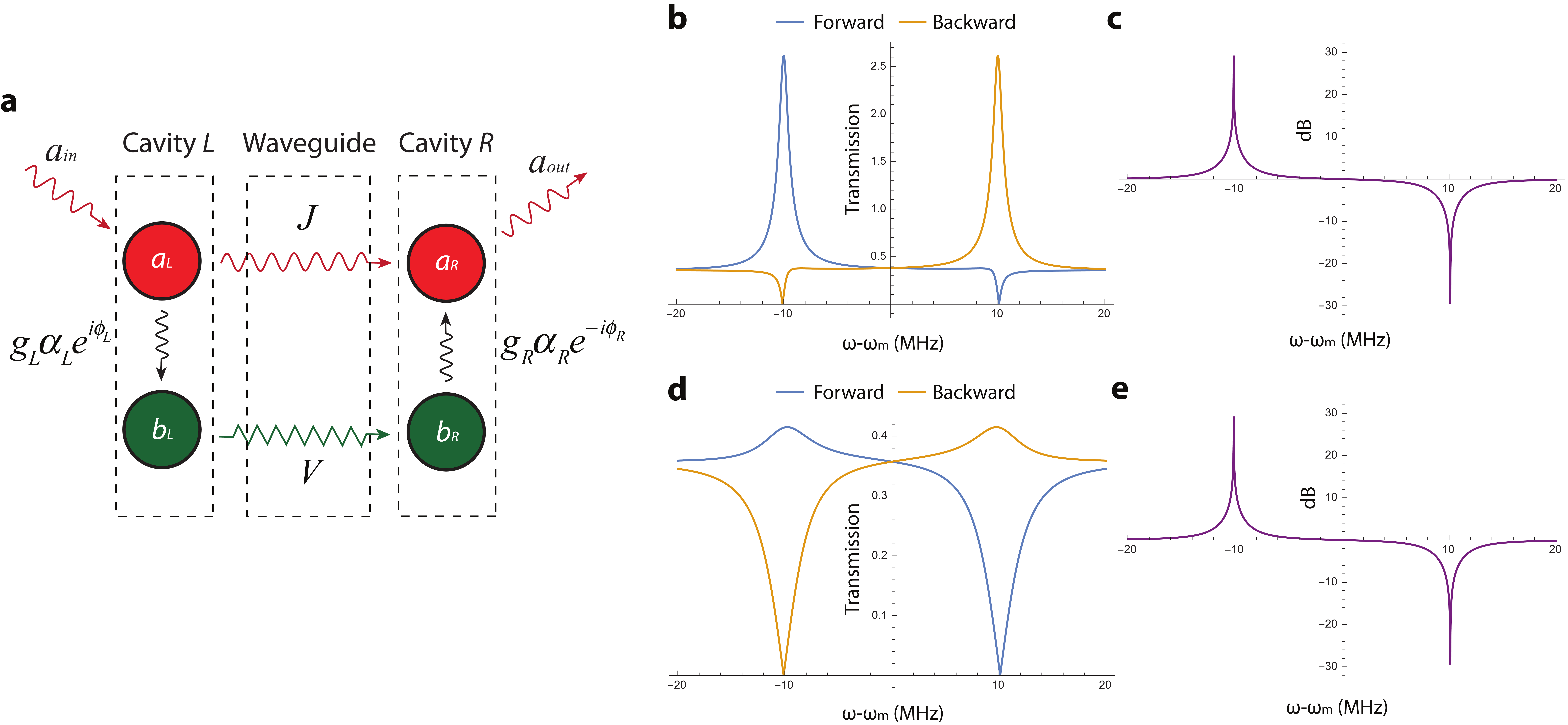}
\caption{\textbf{a} Schematic of the optical non-reciprocity in distantly-coupled optomechanical cavities stemmed from the non-reciprocal optical pump phases and interference between photonic and phononic transmission paths. \textbf{b}-\textbf{e} Transmission coefficient \textbf{b}(\textbf{d}) and isolation ratio \textbf{c}(\textbf{e}) for blue(red) detuned optical pumps with practical device parameters $\gamma_i=2\pi\times 3$ MHz, $g_{L,R}=2\pi\times0.8$ MHz, cavity photon number $n_{cL,R}=2000$,  $\kappa_{L,R}=2\pi\times 2$ GHz, $\kappa_{eL,R}=2\pi\times 1$ GHz, $\phi_L-\phi_R=\frac{\pi}{2}$, $J=2\pi\times 422$ MHz, and $V=2\pi\times 10$ MHz.}
\label{suppfig3}
\end{center}
\end{figure*}

The optical non-reciprocity arisen from this system can be intuitively understood from a schematic shown in Fig. ~\ref{suppfig3}a. The input optical signal undergoes a Mach-Zehnder type of interference through the system: one path is the direct photon hopping and the other path is through radiation-pressure interaction induced transition to phonon and phonon hopping. The phase of the latter path involves the phase difference of the two pumps, which is $\varphi_L-\varphi_R$ for one direction and $\varphi_R-\varphi_L$ for the reversal direction. Such a non-reciprocal phase resembles an effective magnetic flux for photons, resulting in the non-reciprocal transmission~\cite{ab}.

We first consider the case when both cavities are being pumped with blue detuned lasers ($\omega_p-\omega_{ck}=\omega_{mk}$). The equations of motion of the system, in terms of red optical sidebands of the pumps, can be derived from the Hamiltonian of Eq.~\ref{H2} using rotating wave approximation, given sideband resolved condition $\omega_{mk}\gg\kappa_k$, 
\begin{widetext}
\begin{gather}
\label{cme1}\frac{da_L}{dt}=(i\delta_L-\frac{\kappa_L}{2})a_L-iJa_R-ig_L\alpha_Le^{i\phi_L}b_L^\star-\sqrt{\kappa_{eL}}a_{L,{\rm in}},\\
\frac{da_R}{dt}=(i\delta_R-\frac{\kappa_R}{2})a_R-iJa_L-ig_R\alpha_Re^{i\phi_R}b_R^\star-\sqrt{\kappa_{eR}}a_{R,{\rm in}},\\
\frac{db_L}{dt}=-(i\omega_{mL}+\frac{\gamma_i}{2})b_L-iVb_R-ig_L\alpha_Le^{i\phi_L}a_L^\star,\\
\label{cme4}\frac{db_R}{dt}=-(i\omega_{mR}+\frac{\gamma_i}{2})b_R-iVb_L-ig_R\alpha_Re^{i\phi_R}a_R^\star,
\end{gather}
\end{widetext}
where $\delta_k=\omega_{p}-\omega_{ck}$ and $\alpha_ke^{i\phi_k}$ is the steady state amplitude of the local optical cavity mode, which is related to the pumping amplitudes as follows
\begin{widetext}
\be
\alpha_{L(R)}e^{i\phi_{L(R)}}=\frac{(i\delta_{R(L)}-\kappa_{R(L)}/2)\sqrt{\kappa_{eL(R)}}\alpha_{pL(R)}e^{-i\varphi_{L(R)}}+iJ\sqrt{\kappa_{eR(L)}}\alpha_{pR(L)}e^{-i\varphi_{R(L)}}}{(i\delta_L-\kappa_L/2)(i\delta_R-\kappa_R/2)+J^2}.
\ee
\end{widetext}
We find the steady state amplitude is approximately $\sqrt{\kappa_{ek}}\alpha_{pk}e^{-i\varphi_{k}}/i\delta_k$ under the condition $\delta_k\approx\omega_{mk}\gg \kappa_k, J$, which means each cavity is effectively only driven by its own optical pump. This can be intuitively understood by the fact that even and odd hybridized optical cavity modes are driven equally (as $\delta_k\gg J$) and thus the amplitude of one local cavity mode is cancelled out and effectively not being driven. Thus, each cavity-enhanced optomechnical coupling can be independently controlled by the pump.

After solving the equations of motion, we calculate the ratio between right transmission coefficient $T_R$ and left transmission coefficient $T_L$ of the optical signal and find
\be
\frac{T_R}{T_L}=\frac{J-\frac{Vg_Lg_R\alpha_L\alpha_R}{(i(\omega-\omega_{mL})+\frac{\gamma_i}{2})(i(\omega-\omega_{mR})+\frac{\gamma_i}{2})+V^2}e^{ i(\phi_L-\phi_R)}}{J-\frac{Vg_Lg_R\alpha_L\alpha_R}{(i(\omega-\omega_{mL})+\frac{\gamma_i}{2})(i(\omega-\omega_{mR})+\frac{\gamma_i}{2})+V^2}e^{-i(\phi_L-\phi_R)}}.
\ee
Interestingly, this ratio is not explicitly dependent on $\delta_k$ and $\kappa_k$ as an intrinsic property of the device. At the poles $\omega=(\omega_{mL}+\omega_{mR}\pm\sqrt{(\omega_{mL}-\omega_{mR})^2+V^2})/2$, i.e. frequency of the hybridized mechanical modes, and assuming $t\gg \gamma_i$, we have  
\be
\frac{T_R}{T_L}=\frac{J\pm i\frac{Vg_Lg_R\alpha_L\alpha_R}{\gamma_i\sqrt{V^2+(\omega_{mL}-\omega_{mR})^2/4}}e^{ i(\phi_L-\phi_R)}}{J\mp i\frac{Vg_Lg_R\alpha_L\alpha_R}{\gamma_i\sqrt{V^2+(\omega_{mL}-\omega_{mR})^2/4}}e^{-i(\phi_L-\phi_R)}}.
\ee
Thus perfect non-reciprocity, i.e. one direction has vanishing transmission while the other direction has maximal transmission, can be achieved by satisfying the following condition 
\be\label{condition}
\phi_L-\phi_R=\pm\frac{\pi}{2},\quad J=\frac{Vg_Lg_R\alpha_L\alpha_R}{\gamma_i\sqrt{V^2+(\omega_{mL}-\omega_{mR})^2/4}}.
\ee
Under this condition, the transmission coefficient for the through direction is (for simplicity assuming $\omega_{mL}=\omega_{mR}$)
\be\label{maxtransblue}
T_{\neq 0}=\sqrt{\kappa_{eL}\kappa_{eR}}\frac{2\frac{g_Lg_R\alpha_L\alpha_R}{\gamma_i}}{(\frac{\kappa_L}{2}-\frac{g_Lg_R\alpha_L\alpha_R}{\gamma_i})(\frac{\kappa_R}{2}-\frac{g_Lg_R\alpha_L\alpha_R}{\gamma_i})}.
\ee

Similar results can be derived for the case of red detuned pumps, and we find the perfect non-reciprocity condition (Eq.~\ref{condition}) remains the same; while the transmission coefficient for the through direction at poles is given by
\be\label{maxtransred}
T_{\neq 0}=\sqrt{\kappa_{eL}\kappa_{eR}}\frac{2\frac{g_Lg_R\alpha_L\alpha_R}{\gamma_i}}{(\frac{\kappa_L}{2}+\frac{g_Lg_R\alpha_L\alpha_R}{\gamma_i})(\frac{\kappa_R}{2}+\frac{g_Lg_R\alpha_L\alpha_R}{\gamma_i})}.
\ee
We note, in general, an amplified transmission in blue detuned case and an attenuated transmission in red detuned case for the through direction at poles. One can prove from Eq.~\ref{maxtransred} that in the red detuned case, $T_{\neq 0}\le\sqrt{\kappa_{eL}\kappa_{eR}/(\kappa_{L}\kappa_{R})}<1$ and equality is achieved when $\kappa_k/2=\frac{g_Lg_R\alpha_L\alpha_R}{\gamma_i}$. Comparing to Eq.~\ref{condition}, the maximal transmission efficiency is achieved when loss rate $\kappa_k/2$ and coupling rate $J$ is matched at the two cavities (note we used $\omega_{mL}=\omega_{mR}$ for Eqs.~\ref{maxtransblue} and \ref{maxtransred}).  

Based on our fabricated devices, realizing the conditions for perfect non-reciprocity (Eq.~\ref{condition}) is quite promising. For a typical device with $\gamma_i=2\pi\times 3$ MHz, $g_L=g_R=2\pi\times0.8$ MHz, maximal available cavity photon number $n_{cL}=n_{cR}=2000$, and assuming $\omega_{mL}=\omega_{mR}$, Eq.~\ref{condition} determines $J_{\rm max}=422$ MHz. Thus, as long as $J\leq J_{\rm max}$ in this device, perfect non-reciprocity condition can always be achieved by tuning the pump power and phase. We have numerically simulated waveguide (without acoustic shielding) mediated optical coupling between two optomechanical cavities to be 500 MHz and less. Thus it is indeed viable to demonstrate optical non-reciprocity in our devices. Using these parameters along with $\delta_k=\omega_{mk}$, $\kappa_k=2\pi\times 2$ GHz, $\kappa_{ek}=2\pi\times 1$ GHz and $\phi_L-\phi_R=\frac{\pi}{2}$, $J=2\pi\times 422$ MHz, $V=2\pi\times 10$ MHz, we plotted the transmission coefficients and isolation ratio as calculated from Eqs.~\ref{cme1}--\ref{cme4} for the blue-detuned pump In Fig.~\ref{suppfig3}b,c and red-detuned pump In Fig.~\ref{suppfig3}d,e, respectively. 

\end{document}